\documentclass[fleqn,10pt,twocolumn]{SICE14_fixed}
\usepackage{amsmath,amssymb,bbm}
\usepackage{amsfonts}
\usepackage{dsfont}
\usepackage{calligra}
\usepackage{calrsfs}
\usepackage[mathscr]{euscript}

\def\bra#1{\mathinner{\langle{#1}|}}
\def\ket#1{\mathinner{|{#1}\rangle}}

\newcommand{\dbra}[1]{\langle\!\langle{#1}|}
\newcommand{\dket}[1]{|{#1}\rangle\!\rangle}

\def\expect#1{\langle#1\rangle}

\def\cal#1{\mathcal{#1}}
\def\hatcal#1{\hat{\mathcal{#1}}}
\newcommand{\LL}{{\hatcal L}}
\newcommand{\DD}{{\hatcal D}}

\def\bb#1{\mathbf{#1}}
\def\mbb#1{\mathbb{#1}}

\newcommand{\ii}{ {\rm i} }
\newcommand{\dd}{ {\rm d} }
\newcommand{\ZZ}{\mathbb{Z}}

\newcommand{\CC}{\mathbb{C}}

\newcommand{\z}{{\rm z}}

\newcommand{\mm}[1]{{\mathbf{#1}}}
\newcommand{\half}{\frac{1}{2}}

\def\tr{{{\rm tr}}}

\def\ad{{\,{\rm ad}\,}}
\def\End{{\,{\rm End}\,}}
\def\one{{\,{\mathbbm{1}}\,}}

\def\im{{\,{\rm Im}\,}}

\newcommand{\La}{{\mathtt L}}
\newcommand{\Ra}{{\mathtt R}}

\def\pd#1{\partial_{#1}}

\title{Integrable non-equilibrium steady state density operators for boundary driven $XXZ$ spin chains: observables
 and full counting statistics }

\author{Toma\v z Prosen ${}^{1}$ and Berislav Bu\v ca${}^{2}$}

\affils{${}^{1}$Department of Physics, FMF,  University of Ljubljana, Jadranska 19, 1000 Ljubljana, Slovenia\\
(E-mail: tomaz.prosen@fmf.uni-lj.si )\\
${}^{2}$Department of Physics, FMF,  University of Ljubljana, Jadranska 19, 1000 Ljubljana, Slovenia\\
(E-mail: berislav.buca@fmf.uni-lj.si )\\
}
\abstract{%
We will review some known exact solutions for the steady state of the open quantum Heisenberg $XXZ$ spin chain coupled to a pair of baths \cite{p}. The dynamics is modelled by the Lindblad master equation. We also review how to calculate some relevant physical observables and provide the statistics of spin current assuming the spin chain is weakly coupled to the baths \cite{count}.}

\begin{document}

\maketitle


\section{Introduction}

In out-of-equilibrium physics the properties of the long time limit of systems are of paramount importance. Generically, systems when driven out-of-equilibrium return to equilibrium after sufficiently long times. However, cases when they do not are more interesting. There exists various setups which allow this. Here we will focus on systems driven out-of-equilibrium by a pair of infinite baths. The dynamics of the entire system+baths follows the standard Liouville-von Neumann equation,
\begin{equation}
\frac{\dd}{\dd t}\rho(t)=\LL\rho(t)=  -\ii[H_{\rm{sys+env}},\rho(t)], 
\end{equation}
which is impossible to solve. However, since we are only interested in the dynamics of the system with Hamiltonian, $H$, we can trace out the bath part. Then, if we invoke certain approximations, namely the Born-Markov and rotating wave approximation and assume that the coupling between the system and the baths is weak, we are left with the Lindblad master equation for the reduced system dynamics \cite{oqs},
\begin{align}
\frac{\dd}{\dd t}\rho_{\rm{sys}}&=\LL \rho_{\rm{sys}}=-\ii [H,\rho_{\rm{sys}}]+\DD(\rho_{\rm{sys}}),\nonumber \\
\DD(\rho_{\rm{sys}})&:=\sum_{k}\Gamma_{k}\left(L_{k}\rho_{\rm{sys}}L^{\dagger}_{k}-\half\left\{L^{\dagger}_{k}L_{k},\rho_{\rm{sys}}\right\}\right),
\label{lindblad}
\end{align}
where $\DD$ is the non-unitary part of the dynamics representing the action of the baths on the system with so-called Lindblad jump operators  $L_k$, and $\Gamma_k$ are the coupling constants. 
It will be useful to define $(\!\ad H)\rho \equiv [H,\rho]$, where $\!\ad H$ is a linear superoperator acting on the space of operators).
Of particular importance is the fixed point, or steady state, $\rho_{\infty}$, 
\begin{equation}
\LL \rho_{\infty}=-\ii \ad H\rho_{\infty}+\DD(\rho_{\infty})=0. \label{steadystate}
\end{equation}
We will focus on this state for the Heisenberg $XXZ$ spin chain of size $n$ given by the Hamiltonian,
\begin{eqnarray}
H^{\rm{XXZ}}&=&\sum_{j=1}^{n}h^{\rm{XXZ}}_{j,j+1}= \nonumber \\
&=&\sum_{j=1}^{n}2(\sigma^{+}_{j}\sigma^{-}_{j+1}+\sigma^{-}_{j}\sigma^{+}_{j+1})+
\Delta \sigma^{\rm{z}}_{j}\sigma^{\rm{z}}_{j+1}, \label{ham}
\end{eqnarray}
where $\sigma^{\alpha}_{j}$ are the standard Pauli matrices acting on site $j$. Namely,
\begin{equation}
\sigma^{\alpha}_{j}\equiv  \one_{2^{j-1}}\otimes \sigma^{\alpha}\otimes \one_{2^{n-j}}, 
\label{eqn:spin_varibles}
\end{equation}
and $\one_k$ is the identity matrix of dimension $k$. The parameter $\Delta$ is the anisotropy. 

\section{Quantum integrability}
Referring to \cite{iphd} consider the $\calligra{U_{q}}  ({\mathfrak{sl}}_{2}) $ algebra, which is a quantum deformation of the ${\frak{sl}}_{2}$ algebra and is defined by the standard $q-$deformed algebraic relations of its generators, ${\bb{s}}^{\rm{z}}_{s}, {\bb{s}}^{+}_{s}, {\bb{s}}^{-}_{s}$,
 \begin{equation}
[{{\bb{s}}}^{+}_{s},{{\bb{s}}}^{-}_{s}]=[2 {{\bb{s}}}^{\rm{z}}_{s}]_{q},\quad [{\bb{s}}^{\rm{z}}_{s},{\bb{s}}^{\pm}_{s}]=\pm {\bb{s}}^{\pm}_{s}, \label{algebra}
\end{equation}
given by a Verma module representation, ${\frak{S}}_{s}$
 \begin{align}
\label{eqn:hw_spins}
{{\bb{s}}}^{\rm{z}}_{s}&=\sum_{k=0}^{\infty}(s-k)\ket{k}\bra{k},\nonumber \\
{{\bb{s}}}^{+}_{s}&=\sum_{k=0}^{\infty}[k+1]_{q}\ket{k}\bra{k+1},\nonumber \\
{{\bb{s}}}^{-}_{s}&=\sum_{k=0}^{\infty}[2s-k]_{q}\ket{k+1}\bra{k},
\end{align}
and characterised by a continuous spin parameter $s$, where $[x]_q := (q^x-q^{-x})/(q-q^{-1})$. This then allows to define a general $\calligra{U_{q}}  ({\mathfrak{sl}}_{2}) $ invariant Lax operator,  ${\bb{L}}(\varphi,s)\in \End({\frak{S}}_{1/2}\otimes {\frak{S}}_{s})$,
\begin{equation}
{\bb{L}}(\varphi,s)=
\begin{pmatrix}
\sin{(\varphi +\gamma {\bb{s}}^{\rm{z}}_{s})} & (\sin{\gamma}){\bb{s}}^{-}_{s} \cr
(\sin{\gamma}){\bb{s}}^{+}_{s} & \sin{(\varphi -\gamma {\bb{s}}^{\rm{z}}_{s})} 
\end{pmatrix}, \label{lax}
\end{equation}
where $\varphi\in \CC$ is the spectral parameter. Note that ${\frak{S}}_{1/2}$ corresponds to the physical space and ${\frak{S}}_{s}$ to the ``auxiliary space". Defining $\Delta= \rm{cos} (\gamma)$, $q=\rm{exp} (\ii \gamma)$, and, like Eq. \eqref{ham}, $h^{\rm{XXZ}}=2 (\sigma^{+} \otimes_{p} \sigma^{-}+\sigma^{-} \otimes_{p} \sigma^{+})+
\Delta \sigma^{\rm{z}} \otimes_{p} \sigma^{\rm{z}}$ (where $\otimes_{p}$ refers to the tensor product in physical space), one can show that the Sutherland equation holds,
\begin{eqnarray}
&&\!\!\!\![h^{\rm{XXZ}},{{\bb{L}}}(\varphi,s)\otimes_{p}{\bb{L}}(\varphi,s)]= \label{sutherland} \\
&&\!\!\!\!2\sin{\gamma}({\bb{L}}(\varphi,s)\otimes_{p} {\bb{L}}_{\varphi}(\varphi,s)-
{{\bb{L}}}_{\varphi}(\varphi,s)\otimes_{p}{\bb{L}}(\varphi,s)), 
 \nonumber
\end{eqnarray}
with
\begin{align}
{\bb{L}}_{\varphi}(\varphi,s)& := \partial_{\varphi}{\bb{L}}(\varphi,s)\\
& =\cos{\varphi} \cos{(\gamma {\bb{s}}^{\rm{z}}_{s})}\otimes \sigma^{0}-
\sin{\varphi} \sin{(\gamma {\bb{s}}^{\rm{z}}_{s})}\otimes \sigma^{\rm{z}}, \nonumber
\end{align}
which is just the differential (in the spectral parameter $\varphi$) form of the famous Yang-Baxter equation. 

\section{Maximally driven open $XXZ$ spin chain}

Now we will present the results of Ref.~\cite{p} using a more compact form developed e.g. in Ref.~\cite{iphd}. Recall the Lindblad master equation, Eq. \eqref{lindblad} and take two Lindblad jump operator acting on boundary sites $1$ and $n$ of the spin chain, $L_1$ and $L_n$, representing a pure source and pure sink on the two ends,
\begin{equation}
L_{1}=\sqrt{\varepsilon}\sigma^{+}_{1},\qquad L_{n}=\sqrt{\varepsilon}\sigma^{-}_{n},
\label{maxdriv}
\end{equation}
where $\varepsilon$ is the spin bath coupling. 
We are interested in the steady state of the system defined by Eq. \eqref{steadystate}, for which we take a now ubiquitous Cholesky-like ansatz,
\begin{equation}
\rho_{\infty}=S S^{\dagger}, \label{ansatz} 
\end{equation}
with
\begin{equation}
S=\bra{0}{{\bb{L}}}(\varphi,s)^{\otimes n}\ket{0}, \label{sol}
\end{equation}
where ${{\bb{L}}}(\varphi,s)$ is the Lax operator defined by Eq. \eqref{lax}.
Plugging Eq. \eqref{ansatz} into the equation for the steady state Eq. \eqref{steadystate} and using the Sutherland equation Eq. \eqref{sutherland} we are left with 10 boundary equations which we can solve for the spin parameter $s$ and the spectral parameter $\varphi$. The solution is unique and is given by
\begin{align}
\varepsilon&=4\ii[s]^{-1}_{q}\cos{(\gamma s)}, \\
\varphi&=0.
\label{solution}
\end{align}
In the isotropic case $\Delta=1$ ($\gamma \rightarrow 0$) this reduces to,
\begin{equation}
s=4 \ii \varepsilon^{-1}.
\end{equation}
Note that since $s$ is purely imaginary $({\bb{L}}(\varphi, s))^\dagger=({\bb{L}}(\varphi, -s))^{\rm{T}}$, $({\bb{L}}(\varphi,-s))^{\rm{T}}\in \End({\frak{S}}_{1/2}\otimes {\frak{S}}_{-s})$
From now on, since we have calculated $\varphi$ and $s$ we will write for simplicity ${\bb{L}} := {\bb{L}}(\varphi,s)$. 
Since we know the steady state density operator explicitly we can compute the expectation values of observables in the long time limit.

\section{Expectation values of observables}
We are interested in the expectation value of a local observable $O$ in the long time limit,
\begin{equation}
\expect{O}=\frac{\tr ( O \rho_\infty) }{\tr \rho_\infty}
\end{equation}
Now that we know the matrix product state form of the steady state $\rho_\infty$ explicitly, these observables can be computed efficiently numerically for arbitrary system size $n$, as well as, in certain cases, their exact analytical expression in the thermodynamic limit, $n \rightarrow \infty$. \\
First let us define, following Eq. \eqref{ansatz}, a double Lax operator $\mbb{L}$  $\in \End({\frak{S}}_{1/2}\otimes {\frak{S}}_{s}\otimes {\frak{S}}_{-s} )$,
\begin{equation}
{\mbb{L}}={\bb{L}} \otimes_a {\bb{L}}^\dagger,
\end{equation}
where $\otimes_a$ refers to a tensor product of a pair of auxiliary spaces. 

For simplicity we will focus on only the two of the most relevant observables, the spin profile $\expect{\sigma^{\rm{z}}_{k}} $ and spin current, $\expect{j^z_{k}}=\ii\expect{ (\sigma^+_k \sigma^{-}_{k+1}-\sigma^-_k \sigma^{+}_{k+1})}$. 
To calculate these expectation values we will utilise three operators defined using Eq. \eqref{ansatz},
\begin{align}
{\mbb{T}} &=\tr_p \mbb{L} \\
{\mbb{V}} &=\tr_p (\sigma^z \mbb{L}) \\
{\mbb{W}} &=\tr_p( \ii (\sigma^+ \otimes_p \sigma^{-}_{k+1}-\sigma^-  \otimes_p \sigma^{+})  {\mbb{L}} \otimes_p  \mbb{L}),
\end{align}
where the (partial) trace and the outer product are taken with respect to the physical space only. We will write $\ket{0} \otimes_a \ket{0} := \dket{0}$
Furthermore, we should define the trace of the un-normalised steady-state density operator as the non equilibrium partition function, 
\begin{equation}
{\cal{Z}}_{n}=\tr \rho_{\infty}=\dbra{0}{\mbb{T}}^{n}\dket{0}.
\end{equation}
The first useful relation between these operators, which can be shown using the algebraic relations Eq. \eqref{algebra}, and which  manifests the continuity equation for $\sigma^{\rm z}_k$ is,
\begin{equation}
{\mbb{W}}=-2i[s]_{q} \mbb{T},
\end{equation}
which leads to a result first seen in related classical asymmetric exclusion process models \cite{be},
\begin{equation}
\expect{J}:=\expect{j^z_{k}}=-2i[s]_{q} \left(\frac{{\cal{Z}}_{n-1}}{{\cal{Z}}_{n}}\right).
\end{equation}

\begin{figure}
          \centering	
	\includegraphics[width=1.05\columnwidth]{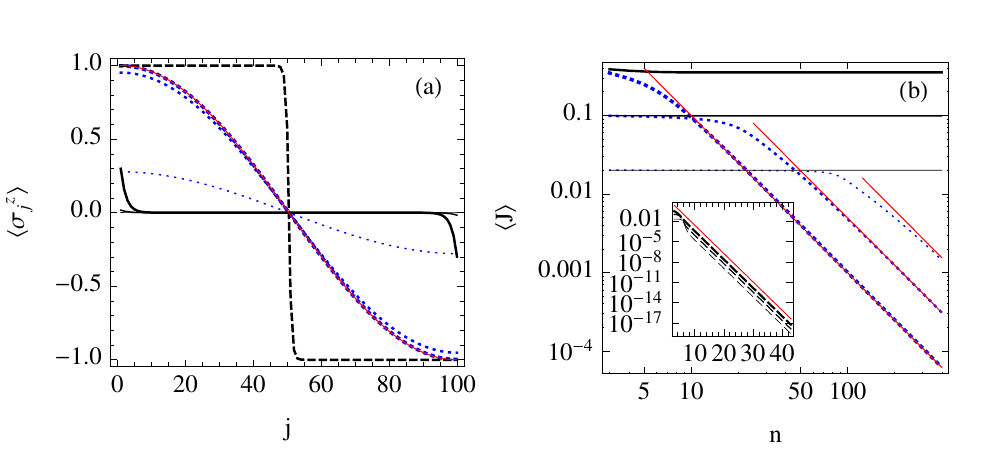}
	\vspace{-6mm}
	
	\caption{(As taken from \cite{p}.)
	Spin profiles $\expect{\sigma^{\rm z}_j}$ at $n=100$ (a), and spin currents $\expect{J}$ vs. size $n$ (b), for $\Delta=3/2$ (dashed), $\Delta=1$ (dotted/blue), 
	$\Delta=1/2$ (full curves), all for three different couplings $\varepsilon=1,1/5, 1/25$ using thick, medium, thin curves, respectively. 
	Red full curves show closed-form asymptotic results  [see text]:
	$\expect{\sigma^{\rm z}_j}=\cos \pi \frac{j-1}{n-1}$, $\expect{J} =\pi^2 \varepsilon^{-1} n^{-2}$ for $\Delta=1$ in the main panels (a,b),
	and  $\expect{J} \propto e^{-n \,{\rm arcosh}\Delta}$ in (b)-inset.}
 	\label{fig:results}
\end{figure}

The operators $\mbb{T}, \mbb{V}$ have effective finite rank $m+1$ for a dense set of anisotropies given by rational numbers $l/m$,
$\gamma = \pi l/m$, (recall that $\Delta= \cos (\gamma)$)  which densely cover the easy-plane regime $|\Delta| < 1$. This allows for closed form expressions for small $m$ where we can diagonalize these operators analytically. For example, for $\Delta = 1/2=\cos \pi/3$, in the thermodynamic limit, one finds flat spin profiles and can prove ballistic transport by explicitly computing the limit (see Fig. \ref{fig:results}) 
$\expect{J}|_{n\to\infty} =  \frac{\left(\sqrt{ 81 + 74 \varepsilon^2 + 9 \varepsilon^4} - 7 - 3 \varepsilon^2 \right)\varepsilon}{4(1+\varepsilon^2)}$.

In contrast, in the the easy-axis regime $|\Delta|>1$, where the before-used effective truncation of the auxiliary space is not possible, one finds an asymptotically exponentially decaying current and hence insulating behaviour, $\expect{J} \propto (|\Delta| + \sqrt{\Delta^2-1})^{-n}$ (Fig.~\ref{fig:results}b-inset) with consistent kink-shaped magnetization profiles (Fig.~\ref{fig:results}a). 

Finally let us examine the critical isotropic point $\Delta=1$, $s=\frac{4\ii}{\varepsilon}$. We will utilise a useful algebraic relation between $\mbb{T}$ and $\mbb{V}$,
\begin{equation}
[{\mbb{T}},[{\mbb{T}},{\mbb{V}}]]+2\{{\mbb{T}},{\mbb{V}}\}-8s^{2}{\mbb{V}}=0. \label{vtalg}
\end{equation}
We will also need, 
\begin{equation}
\frac{{\cal{Z}}_n}{{\cal{Z}}_{n-1}} \simeq \varepsilon^2 \biggl({ \frac{(4n-3)^2}{32\pi^2}} - \alpha\biggr)+1+ {\cal O}(n^{-1}), \label{asymT}
\end{equation}
where $\alpha$ is a constant,
and 
the boundary conditions,
\begin{equation}
\dbra{0}({\mbb{T}}-{\mbb{V}})=\dbra{0},\quad ({\mbb{T}}+{\mbb{V}})\dket{0}=\dket{0}. \label{bound}
\end{equation}
Then we can multiply the algebraic relation Eq. \eqref{vtalg}  by $\dbra{0}{\mbb{T}}^{j-1}$ from the left, and ${\mbb{T}}^{n-j-2}\dket{0}$ from the right and use Eq. \eqref{asymT} and the boundary equations from Eq. \eqref{bound} and in the continuum limit, $M (x\equiv \frac{j-1}{n-1}):=\expect{\sigma^{\rm z}_j}$, find a second-order differential equation $M''(x)=- \pi^2 M(x) + {\cal O}(\frac{1}{n})$ and solve with $M(x) = \cos \pi x$, or $\expect{\sigma^{\rm z}_j} \simeq \cos \pi \frac{j-1}{n-1}$. 
Also, Eq.\eqref{asymT} implies anomalous sub-diffusive scaling $\expect{J}\approx \pi^2 \varepsilon^{-1} n^{-2}$ to leading order in  the thermodynamic limit.

\section{Full counting statistics}

In this section we present the results from  Ref.~\cite{count}. Another interesting question concerning observables, with them being probabilistic in nature, is what are, not only their expectation values, but all other higher moments as well? The most relevant physical observable out-of-equilibrium is the current of some quantity. It is very challenging to study these fluctuation properties of the current which in general depend not only on the (asymptotic) steady state of the system, but also on the correlations at earlier times. Therefore, our knowledge of the steady state density operator is not enough to compute the higher moments of spin current, $J$, even though it is enough to compute just $\expect{J}$, as shown in the previous section. 

Spin current is the flow of magnetization, $M:=\sum_{j=1}^n \sigma^{\rm{z}}_j$. That is, if we label the amount $M$ transported in time $t$ from the first bath to the second bath by $N(t)$ then in the long time limit, $J=\lim_{t\to \infty}\frac{1}{2 t} N(t)$.

Magnetization is conserved by the Hamiltonian, but is changed by the Lindblad jump operators, Eq. \eqref{lindblad}.  Therefore, the Lindblad jump operators are the ones which drive the current through the system. One approach to this problem, using master equations, which has attracted a lot of attention quite recently \cite{z1,z2,z3,mh,mk1,mk2}, is the method of so-called ``full counting statistics" by utilising a ``counting field". In order to use this method first let us split the dissipator in the Lindblad master equation Eq. \eqref{lindblad} into a jump part and a dissipative part, 
\begin{equation}
\DD^{\rm jump}\rho :=\sum_\mu L_m \rho L^\dagger_m,\quad
\DD^{\rm diss} \rho := \frac{1}{2}\sum_m \{ L^\dagger_m L_m,\rho\},
\end{equation}
Depending on whether $L_m$ changes magnetization $M$ by $\pm 1$, i.e., $[M,L_m] = \pm L_m$, one modifies the jump part of the dissipator $\DD^{\rm jump}$ by introducing a counting field, $\chi$, 
\begin{equation}
\DD_\chi^{\rm jump}\rho :=\sum_{m} e^{ \pm \ii \chi} L_{m} \rho L^\dagger_{m}
\end{equation}
which counts how many times a certain Lindblad jump operator has driven the current in either the left (+) or right (-) direction. 
With this modification in place one can show that the cumulants of the current are given by,
\begin{equation}
\expect{J^m}_c := \lim_{t \to \infty}\frac{1}{2 t} \expect{[N(t)]^m}_c =  \frac {\partial^{m} \lambda(\chi) }{\partial (\ii \chi)^{m}} \Big |_{\chi \to 0}. \label{cs0}
\end{equation}
Here $\lambda(\chi)$ is a leading eigenvalue (of maximal real part) of the modified Liouvillean,

\begin{equation}
\left [-\ii \ad H + \varepsilon \left ( \DD_\chi^{\rm jump} + \DD^{\rm diss} \right ) \right ]\rho(\chi) = \lambda(\chi)\rho(\chi) 
\label{eq:gen}
\end{equation}
The full proof of this statement is rather long \cite{rev}, however it may be intuitively understood by observing a reduced density matrix $\rho_N(t)$, that is $\rho(t)$ projected to a subspace of $N$ spin-transfers between the two baths in time $t$. The trace of this, $P_N(t)=\tr \rho_N(t)$, is the probability of $N$ spin transfers in time $t$. By performing a Fourier transform (in $N)$ of this reduced density matrix, $\rho( \chi, t)=\sum_N \rho_N(t) e^{-\ii \chi N}$, it may then be shown (by observing the action of the generator of the time evolution) that the Lindblad master equation, Eq. \eqref {lindblad}, has the jump superoperator  modified, $\DD_\chi^{\rm jump}\rho :=\sum_{\mu,\nu} e^{\ii \mu \nu \chi} L_{\mu, \nu} \rho L^\dagger_{\mu,\nu}$, so that it depends on the counting field $\chi$ and in which direction ($\mu \nu $) a specific Lindblad operator $L_{\mu,\nu}$ drives the flow. Furthermore,  if we normalize $\tr \rho (\chi, t=0)=1$, the largest eigenvalue of the Liouvillian, $\lambda (\chi)$, corresponds to the cumulant generating function for the current distribution in the long time limit \cite{rev}, since for large $t$, $\rho(\chi, t) \approx e^{\lambda(\chi) t} \rho(\chi,t= 0) $. 

When we take $H=H^{\rm{XXZ}}$, Eq. \eqref{ham} and maximal driving driving, Eq. \eqref{maxdriv} finding $\lambda(\chi)$ analytically seems impossible. 
However, we are in good shape since one of the assumptions when deriving the Lindblad master equation is that the system-bath coupling $\varepsilon$ is weak. This allows us to perform a perturbation series expansion in $\varepsilon$. 
Quite remarkably, when we do this we find universal results, in the leading order for the full counting statistics of spin current, not just for $H^{\rm{XXZ}}$, but for any parity symmetric Hamiltonian, $H$. This condition means that there exists an operator $P$, with $P^2=\one$, such that $P H = H P$, and satisfying at least one of the following properties
\begin{equation}
 \quad P \sigma^\z_1 = \sigma^\z_n P,\quad {\rm or}\quad  P \sigma^\z_{1,n} = -\sigma^\z_{1,n} P, \label{eq:Psym2}
\end{equation}
with an additional {\em weak} requirement, namely that also all conserved operators, $Q_k$ ($\ad H Q_k=0$), are parity symmetric $[P,Q_k]=0$. 
We can also drop the requirement of maximal driving via Lindblad jump operators, Eq. \eqref{maxdriv} and instead use the most general set of Lindblad jump operators acting on site $1$ and $n$,
\begin{align}
L_{+,+} &= \sqrt{ \varepsilon a} \sigma^+_1, \quad L_{+,-} = \sqrt{ \varepsilon b} \sigma^-_1, \quad \nonumber \\
L_{-,+} &= \sqrt{ \varepsilon c} \sigma^+_n, \quad L_{-,-} = \sqrt{ \varepsilon d} \sigma^-_n.  \label{gendriv}
\end{align}
 See Fig. \ref{fig:CF} for a graphic illustration.
\begin{figure}
 \centering	
\vspace{-1mm}
\includegraphics[width=0.95\columnwidth]{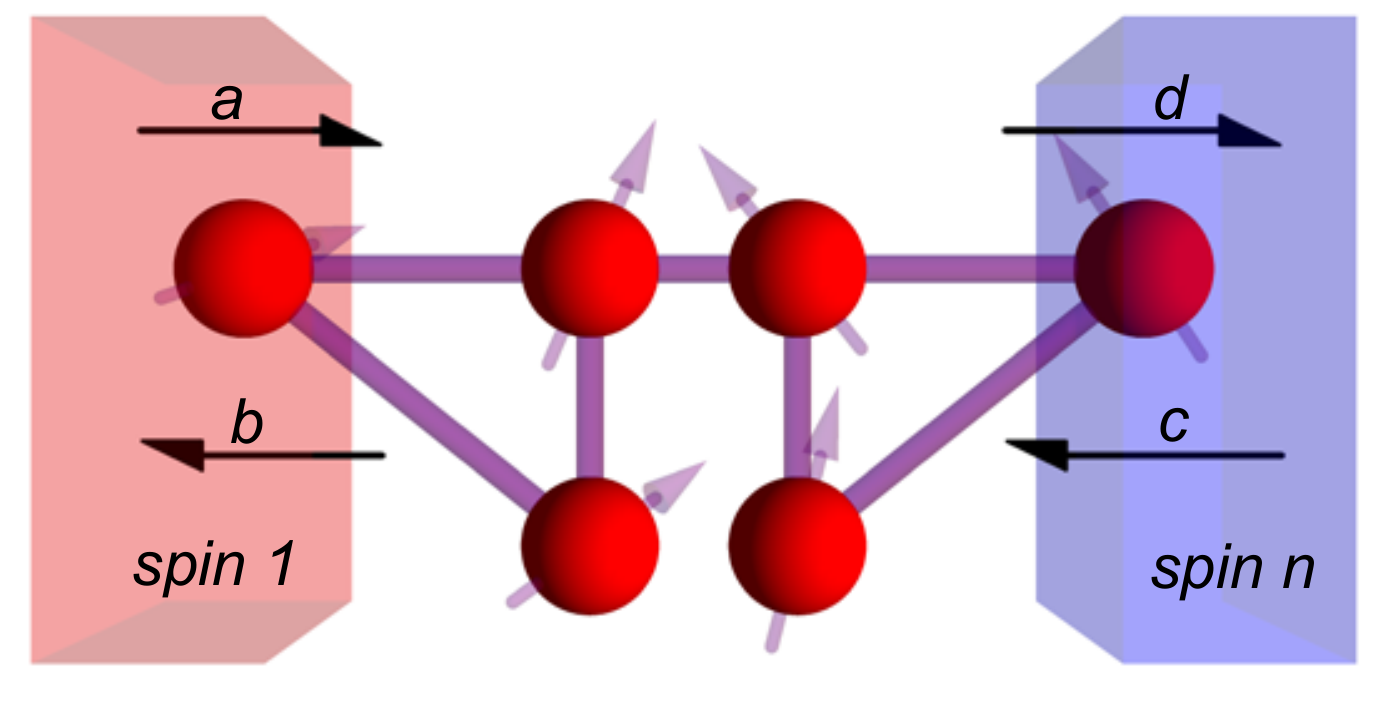}
\vspace{-1mm}
\caption{(As taken from \cite{count}) An illustrative example of a parity symmetric spin system. The system is coupled to the two baths (represented by boxes), at site 1 and site $n$. The baths act via decoherent jump operators with corresponding rates ($a,b,c,d$). The arrows beneath the jump rates  indicate in which direction the spin current is being driven.  }
\label{fig:CF}
\end{figure}
We will assume that the conditions of Evans theorem hold (or the Liouvillean can be symmetry reduced \cite{evans}) so the fixed point $\rho(\chi)$ is unique. We perform a perturbation series in $\varepsilon$ (using the fact that $\lambda(\chi)$ is an odd function of $\varepsilon$),
\begin{equation}
\rho (\chi) = \sum_{p=0}^\infty (\ii \varepsilon)^p \rho^{(p)}(\chi), \; \lambda (\chi)  = \sum_{p=1}^\infty \varepsilon^{2p-1} \lambda^{(2p-1)}(\chi),
\end{equation}
Plugging this into Eq. \eqref{eq:gen} (defining $ \hat{\cal D}_\chi :=\DD_\chi^{\rm jump} + \DD^{\rm diss} $) one arrives to two equations for the first two orders in $\varepsilon$,
\begin{align} (\!\ad H) \rho^{(0)} &=0, \label{zo} \\ 
(\!\ad H) \rho^{(1)} + \hat{\cal D}_\chi \rho^{(0)}  &= \lambda^{(1)} \rho^{(0)}. \label{fo}
\end{align}
For Eq. \eqref{zo} we take as an ansatz,
\begin{equation}
\rho^{(0)} = 2^{-n}\prod_{j=1}^n(\one + \nu(\chi) \sigma_j^{\rm z}),
\label{eq:ansatz}
\end{equation}
with free parameter $\nu(\chi)$, which may be found by requiring that $\hat{\cal D}_\chi \rho^{(0)} - \lambda^{(1)} \rho^{(0)} \in \im (\ad H)$ (that is, there exists a solution to the first order equation Eq. \eqref{fo}), assuming the aforementioned parity symmetry requirement holds, Eq. \eqref{eq:Psym2}.  By taking the trace of the first order equation \eqref{fo} we also find $ \lambda^{(1)} $,
\begin{eqnarray}
\lambda^{(1)}\!\!\!\!&=&\!2\sqrt{1+ (e^{-2 \ii \chi}-1)ad + (e^{2 \ii \chi}-1)bc}  - 2,\; \label{foa} \\
\nu\!&=&\!\frac{\lambda^{(1)}-(e^{-\ii \chi}-1)(a+d)+(e^{\ii\chi}-1)(b+c)}{(e^{-\ii\chi}-1)(a-d)+(e^{\ii\chi}-1)(b-c)}, \nonumber
\end{eqnarray}
 We then apply Eq.~\eqref{foa} to calculate all the cumulants for this wide class of spin systems via Eq.~\eqref{cs0}. For instance, the expectation value of the spin current is $ \left <I_{(1)} \right >_c= \frac{\varepsilon}{2} (a d-b c)$.   Closed form expressions for higher cumulants were obtained in the same way, but are lengthy and therefore we will not write them out. However, they significantly simplify if we consider a symmetric driving instead of a general one \eqref{gendriv}, 
\begin{equation}
a=d=(1+\mu)/2, \qquad b=c=(1-\mu)/2,
\label{eq:symm}
\end{equation}
where the driving strength parameter $\mu$ controls the non-equilibrium forcing due to unequal average spin polarizations of the two baths. Then we have $\nu=0$, $\rho^{(0)}=2^{-n} \one$, so $\lambda^{(1)}=-1+\cos\chi - \ii \mu \sin\chi $, and 
$\expect{I^{2k+1}_{(1)}}_c=\varepsilon \mu/2$ for odd cumulants and $\expect{I^{2k}_{(1)}}_c=\varepsilon/2$ for even cumulants. Extreme driving $\mu=1$ hence results in the Poisson distribution $\expect{I^m}_c={\rm const}$. 

Under symmetric driving, Eq. \eqref{eq:symm}, we may also find the third order correction to the current cumulant generating function $\lambda^{(3)}$ for the $XXZ$ spin chain $H= H^{\rm{XXZ}}$. In order to do this we will turn again to the known solution for maximum driving without the counting field, $\rho_\infty=S S^\dagger$, where $S$ is defined by Eq. \eqref{sol}.

Define an operator $Z$ from $S$ by (see \cite{ip})
\begin{equation}
\pd{s} S |_{s=0}=  \frac{2 \gamma \sin \gamma}{\sin \varphi} Z + \gamma \cos \varphi M,
\end{equation}
where, as before $\Delta=\cos(\gamma)$, $\varphi$ is the spectral parameter, $s$ is the continuous spin parameter and $M$ is the magnetization. 
Calculating $\pd{s} \ad H^{\rm{XXZ}} S |_{s=0}$ using the Sutherland equation Eq. \eqref{sutherland} one finds for $Z$,
\begin{equation}
[H^{\rm{XXZ}},Z]=\sigma^{\rm{z}}_1-\sigma^{\rm{z}}_n. \label{zcom}
\end{equation}
Remembering that for symmetric driving, Eq. \eqref{eq:symm} $\rho_{(0)}=2^{-n} \one $ we can immediately solve the first order equation Eq. \eqref{fo},
$ 2^n \rho^{(1)}=c^{(1)} (Z-Z^\dagger)$ (in a similar way as when there is no counting field $\chi=0$ \cite{p11a}, but multiplied by a different constant), $c^{(1)}= (-\mu-\mu \cos\chi + \ii \sin\chi)/2 $.
The second order equation,
\begin{equation}
 (\!\ad H)\rho^{(2)} + \hat{\cal D}_\chi \rho^{(1)}= \lambda^{(1)} \rho^{(1)}.
 \label{eq:seco}
\end{equation}
 may be solved through a longer computation by taking an ansatz, { similar in form} to the one for $\chi=0$ \cite{p11a}, namely, 
\begin{equation}
2^n \rho^{(2)} = c^{(1)} c^{(2)}_1 (Z - Z^\dagger)^2 - c^{(1)}c^{(2)}_2 [Z,Z^\dagger],
\end{equation}
and finding that $c^{(2)}_1=\frac{1}{4}  (-\mu-\mu \cos\chi + \ii \sin\chi) $ and $c^{(2)}_2=\frac{1}{2} ( \cos\chi - \ii \mu \sin\chi)$.
The solution to the second order equation determines $\rho^{(2)}$ only up to an addition of a linear combination of conserved quantities $Q_k$, $\ad H^{\rm{XXZ}} Q_k$, i.e., the full solution can be written as $\rho^{(2)'}=\rho^{(2)}+\sum_k \alpha_k Q_k$.
These turn out to be irrelevant however, because by taking the trace of the third order equation,
 \begin{equation}
 \lambda^{(3)} = \tr (\hat{\cal D}_\chi \rho^{(2)'}- \lambda^{(1)} \rho^{(2)'}). \label{tr3}
\end{equation}
and noting that the only relevant component of $ \rho^{(2)'}$, $\hat{\cal D}_\chi (\sigma^{\rm{z}}_1-\sigma^{\rm{z}}_n)$, which contributes to the trace, is not a conserved quantity of $\ad H^{\rm{XXZ}}$ due to the existence of a solution $Z$ to Eq. \eqref{zcom}.  

We therefore have,
 \begin{equation}
 \lambda^{(3)} = (-\mu+\mu \cos\chi + \ii \sin\chi)\tr[\bigl(-\sigma^z_1+\sigma^z_n) \rho^{(2)}\bigr] \label{tr3a}.
\end{equation}
This finally allows us to write the third-order correct to the current cumulants,
\begin{eqnarray}
\left <I^{2k}_{(3)} \right >_c &=-\varepsilon^3 f(n) \frac{(9^k-1) (3\mu^2+1)}{128 (2k)!},\quad \\
\left <I^{2k+1}_{(3)} \right >_c &= -\varepsilon^3 f(n)  \frac{ \mu ( 9^{k+1}-1 + 3 (9^k-1) \mu^2)}{256 (2k+1)!} \nonumber
\end{eqnarray}
where $f(n)={\bra{\La} \mm{T}^{n} \ket{\Ra}}-{\bra{\La} \mm{T}^{n-1} \ket{\Ra}}$ and $\mm{T}$ is exactly the transfer matrix from Ref. \cite{p11a}, acting on auxiliary Hilbert space with two ground-state vectors $\ket{\Ra}$ and $\ket{\La}$ (for details of how to derive this transfer matrix see also \cite{ip}). We can again compute $f(n)$ efficiently for any anisotropy $\Delta$ of the form $\Delta = \cos(\pi l/ m)$, $l,m\in\ZZ$. For instance, for $\Delta=1$, we have, $f(n)=n-1$, and for $\Delta=1/2$, $f(n)=\frac{1}{45} (-1)^{-n} 8^{1-n} \left(5 (-8)^n-6 (-5)^n+10\right)$. Finally we contrast our results with a spin system which does not fulfil the parity symmetry requirement from Eq. \eqref{eq:Psym2}, namely a $XXZ$ spin chain in a staggered field as shown in Fig. \ref{fig:CF}
\begin{figure}
 \centering	
\vspace{-1mm}
\includegraphics[width=0.85\columnwidth]{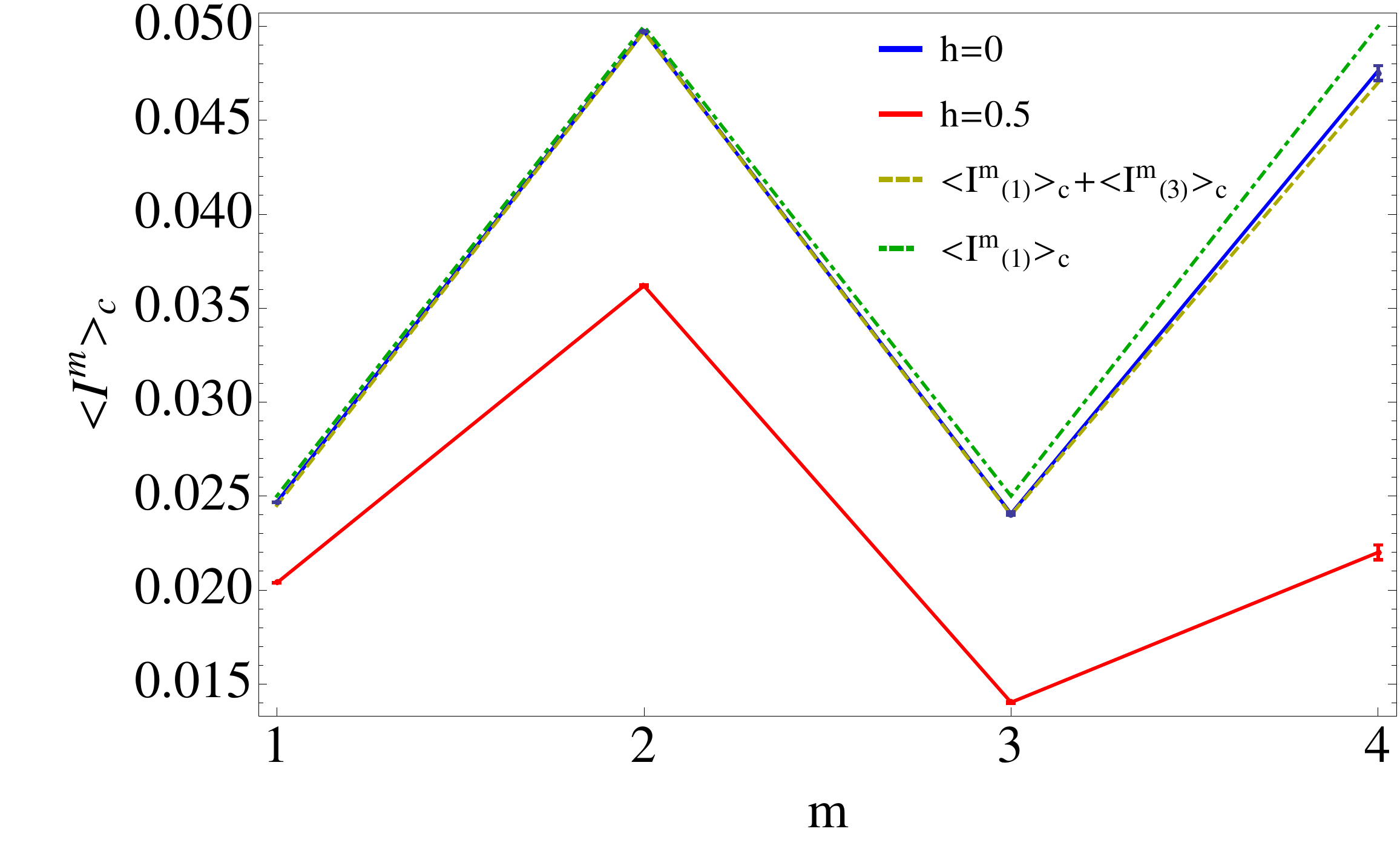}
\vspace{-1mm}
\caption{(As taken from \cite{count}) The first four current cumulants obtained numerically for the $XXZ$ spin chain and staggered field $XXZ$ spin model, $H=H_{XXZ}+ \sum_{j=1}^{n}h  (\pm 1)^j  \sigma_j{\rm{z}}$, with field strength $h$, for $n=4$, $\Delta=0.5$, $\mu=0.5$, $\varepsilon=0.1$.
Dashed (chained) lines indicate analytical results up to second (fourth) order in $\varepsilon$. }
\label{fig:CF}
\end{figure}

\section{Conclusions}
We have shown how to find the steady state analytically for the Heisenberg XXZ spin chain undergoing maximum driving under the Lindblad master equation and compute the spin current, $\langle J\rangle$ and magnetization profiles $\langle\sigma^{\rm{z}}_k\rangle$. Depending on the value of the anisotropy $\Delta$ we find, in the thermodynamic limit $n \rightarrow \infty$, either ballistic (for a dense subset of $| \Delta |<1$), insulating (for $| \Delta |>1$), or anomalous transport ($\Delta=1$) with the spin current $J$ decaying as $J\propto 1/n^2$.
We have also shown what the universal results in leading order of perturbation in system-bath coupling $\varepsilon$ for the full counting statistics of spin current for any driving acting on the boundary sites $1$ and $n$ only are, as well as the next-to-leading order corrections for the Heisenberg XXZ spin chain undergoing symmetric driving.


\end{document}